\documentclass[twocolumn,pra,showpacs,aps]{revtex4-1}
\usepackage{amssymb}
\usepackage{amsmath}
\usepackage{graphicx}
\usepackage{epsfig}
\usepackage{float}

\begin{document}

\title{Coalescing Majorana edge modes in non-Hermitian $\mathcal{PT}$%
-symmetric Kitaev chain}
\author{C. Li, L. Jin and Z. Song}
\email{songtc@nankai.edu.cn}
\affiliation{School of Physics, Nankai University, Tianjin 300071, China}

\begin{abstract}
A single unit cell contains all the information about the bulk system,
including the topological feature. The topological invariant can be
extracted from a finite system, which consists of several unit cells under
certain environment, such as a non-Hermitian external field. We investigate
a non-Hermitian finite-size Kitaev chain with $\mathcal{PT}$-symmetric
chemical potentials. Exact solution at the symmetric point shows that
Majorana edge modes can emerge as the coalescing states at exceptional
points and $\mathcal{PT}$ symmetry breaking states. The coalescing zero mode
is the finite-size projection of the conventional degenerate zero modes in a
Hermitian infinite system with the open boundary condition. It indicates a
variant of the bulk-edge correspondence: The number of Majorana edge modes
in a finite non-Hermitian system can be the topological invariant to
identify the topological phase of the corresponding bulk Hermitian system.
\end{abstract}

\maketitle





\section{Introduction}

The discovery of topological matter which exhibits topological properties in
the band structure has opened a growing research field \cite%
{Hasan,XLQ,CKC,HMW,Lecture notes}. A particularly important concept is the
bulk-edge correspondence, which indicates that, a nontrivial topological
invariant in the bulk indicates localized edge modes that only appear in the
presence of the open boundary in the thermodynamic limit. In general, a bulk
system is constructed by stacking a great many copies of a single original
unit cell in an array. In this sense, a single unit cell contains all the
information about the bulk system, including the topological feature. Then
the topological invariant can be, in principle, extracted from a finite
system, which consists of several unit cells under certain environment, such
as a non-Hermitian external field. In fact, the original bulk-edge
correspondence\ is an example in this context: The open boundary can be
technically regarded as one of the extreme cases of adding local impurity on
the bulk, that breaks the translational symmetry. On the other hand, it has
been shown that a finite system with imaginary ending potentials can share
the common eigenstates with an infinite system \cite{JL 81 83 44}. It
implies that a finite non-Hermitian system may retain some of
characteristics, such as zero-energy modes of an infinite Hermitian system.
An interesting question is whether there is a generalization of the
bulk-edge correspondence to non-Hermitian systems which arises from the
imaginary impurity.

In this work, we investigate a non-Hermitian finite-size Kitaev chain with
parity-time ($\mathcal{PT)}$ symmetric chemical potentials. We demonstrate
that the key to retrieve the Majorana zero mode from a small system is a
pair of specific imaginary chemical potentials, under which the coalescing
zero mode shares the identical pattern with that of the conventional zero
mode in the thermodynamic limit. The coalescing zero mode in a finite-size
non-Hermitian system can be directly obtained from the projection of the
conventional degenerate zero modes in a Hermitian infinite system. Exact
solution also shows the existence of Majorana edge modes, which emerge as a
pair of $\mathcal{PT}$ symmetry breaking states with imaginary eigenvalues.
It indicates a variant of the bulk-edge correspondence: The number of
Majorana edge modes\ in a finite-size non-Hermitian system can be the
topological invariant to identify the topological phase of the corresponding
bulk Hermitian system.

The remainder of this paper is organized as follows. In Sec. \ref{Model
Hamiltonians}, we present the Hamiltonian of the Kitaev ring with two
impurities. In Sec. \ref{Majorana representation} the corresponding Majorana
representation of the model with $\mathcal{PT}$-symmetric chemical
potentials is given. Sec. \ref{Edge modes} devotes to the investigation of
Majorana bound states at the symmetric point. In Sec. \ref{Connection to
conventional zero mode}, we reveal the link between the coalescing zero mode
and the Hermitian one. Finally, we present a summary and discussion in Sec. %
\ref{Summary}.

\begin{figure}[tbp]
\includegraphics[ bb=38 247 563 772, width=0.45\textwidth, clip]{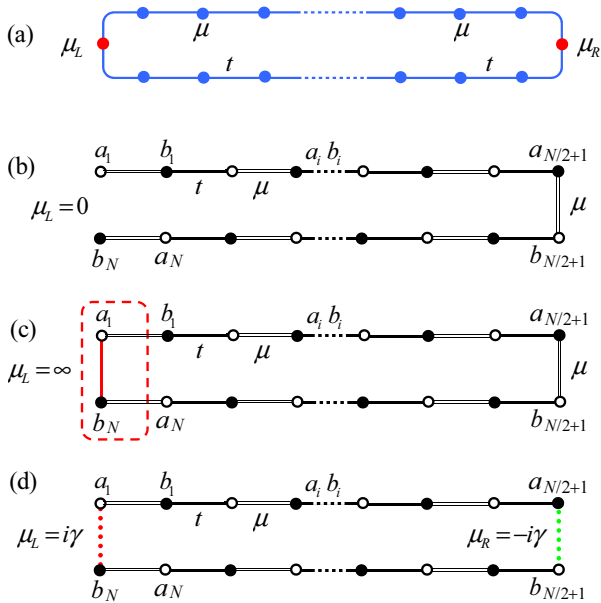}
\caption{(Color online) Schematic illustrations for edge modes in the
Majorana Hamiltonian from Eq. (\protect\ref{H_main}), induced by impurities
embedded in a Kitaev ring. (a) $N$-site Kitaev ring with uniform chemical
potential $\protect\mu$ and hopping amplitude (pairing amplitude) $t$. Two
impurities are located at sites $1$ and $N/2+1$, in terms of specific values
of chemical potentials $\protect\mu_{L}$ and $\protect\mu_{R}$,
respectively. The corresponding Majorana lattice is a $2N$-site dimerized
ring with staggered hopping strengths $t$ (single line) and $\protect\mu$
(double line), which contains two specific dimers with hopping amplitudes $\protect\mu_{L}$
and $\protect\mu_{R}$. We focus on three cases: (b)$\protect\mu_{L}=0$,
$\protect\mu_{R}=\protect\mu$, the lattice becomes a standard $2N $-site SSH chain,
which possesses edge modes for $\protect\mu>1$. (c) $\protect\mu_{L}=\infty$,
$\protect\mu_{R}=\protect\mu$, the dimer (circled
by the dashed red line) is adiabatic eliminated, and then the lattice
becomes a standard $2(N-1)$ SSH chain, which possesses edge modes for $\protect\mu<1$.
For infinite $N$, the conclusions for cases (b) and (c) can
be regarded as the extension of original bulk-edge correspondence.
(d) $\protect\mu_{L}=- \protect\mu_{R}=i\protect\gamma$, it is $\mathcal{PT}$ symmetric.
For finite $N$, it has two coalescing zero modes for $\protect\mu>1$, while
two degenerate zero modes and two imaginary-energy modes for $\protect\mu<1$.
All these modes exhibit evanescent wave behavior. The number of edge modes
for finite system matches the topological invariant of the bulk. It
indicates that bulk-edge correspondence can be extended to the non-Hermitian
regime. } \label{fig1}
\end{figure}

\section{Model Hamiltonians}

\label{Model Hamiltonians}

We consider a one-dimensional Kitaev model with two impurities. The
Hamiltonian of the tight-binding model takes the following form%
\begin{eqnarray}
H &=&H_{0}+H_{\mathrm{im}},  \notag \\
H_{0} &=&-\sum\limits_{j=1}^{N}(tc_{j}^{\dag }c_{j+1}+\Delta c_{j}^{\dag
}c_{j+1}^{\dag }+\text{\textrm{H.c.}})-\mu \sum\limits_{j=1}^{N}\left(
1-2n_{j}\right) ,  \notag \\
H_{\mathrm{im}} &=&(\mu -\mu _{\mathrm{L}})\left( 1-2n_{1}\right) +(\mu -\mu
_{\mathrm{R}})\left( 1-2n_{N/2+1}\right) ,  \label{H_main}
\end{eqnarray}%
where $j$ is the coordinate of lattice sites and $c_{j}$ is the fermion
annihilation operator at site $j$. $H_{0}$ is employed to depict $p$-wave
superconductors.\ The hopping between (pair operator of) neighboring sites
is described by the hopping amplitude $t$ (the real order parameter $\Delta $%
). The last term in $H_{0}$\ gives the chemical potential. Imposing the
periodic boundary condition $c_{1}\equiv c_{N+1}$,\ the Hamiltonian $H_{0}$
can be exactly diagonalized and the topological invariant can be obtained in
various parameter regions. It provides well-known example of systems with
the bulk-edge correspondence when the open boundary condition is imposed. It
turns out that a sufficient long chain has Majorana modes at its two ends
\cite{AYK}. A number of experimental realizations of $p$-wave
superconductors have found evidence for such Majorana modes \cite%
{VM,LPR,AD,ADKF,BA}.\textbf{\ }Term $H_{\mathrm{im}}$\ represents two
impurities on the two symmetrical sites. The impurities are described in
terms of specific values of chemical potentials $\mu _{\mathrm{L}}$ and $\mu
_{\mathrm{R}}$. We note that taking $\mu _{\mathrm{L}}=0$ or $\pm \infty $,
and $\mu _{\mathrm{R}}=\mu $,\ it corresponds to the open Majorana chains
with different edges (see Fig. \ref{fig1}). It is an alternative
representation of bulk-boundary correspondence. In this context, the real
value of $\mu _{\mathrm{L,R}}$\ requires an infinite system. In this work,
we consider imaginary value of $\mu _{\mathrm{L,R}}$. In contrast to
previous studies based on Hermitian chains in the thermodynamic limit, we
focus on the Kitaev model on a finite lattice system. This is motivated by
the desire to get a clear physical picture of the edge mode through the
investigation of a small system. We first study the present model from the
description in terms of Majorana fermions.\ We will show that the evident
indicator for phase diagram does not require infinite system.

\section{Majorana representation}

\label{Majorana representation} In the following, the impurity is taken as $%
\mathcal{PT}$ symmetric with $\mu _{\mathrm{L}}=i\gamma $ and $\mu _{\mathrm{%
R}}=-i\gamma $. Note that the Hamiltonian $H$ is not $\mathcal{PT}$
symmetric unless $\Delta $ is pure imaginary. Here the space-reflection
operator, or the parity operator $\mathcal{P}$\ and the time-reversal
operator $\mathcal{T}$, are defined as%
\begin{equation}
\mathcal{P}c_{l}^{\dagger }\mathcal{P}=c_{N+1-l}^{\dagger }\text{, }\mathcal{%
T}i\mathcal{T}=-i\text{.}
\end{equation}%
Nevertheless, we will see that the core matrix in the Majorana fermion
representation is $\mathcal{PT}$ symmetric, which ensures the system is
pseudo-Hermitian.

We introduce Majorana fermion operators%
\begin{equation}
a_{j}=c_{j}^{\dagger }+c_{j},b_{j}=-i\left( c_{j}^{\dagger }-c_{j}\right) ,
\end{equation}%
which satisfy the relations%
\begin{eqnarray}
\{a_{j},a_{j^{^{\prime }}}\} &=&2\delta _{jj^{^{\prime
}}},\{b_{j},b_{j^{^{\prime }}}\}=2\delta _{jj^{^{\prime }}}  \notag \\
\{a_{j},b_{j^{^{\prime }}}\} &=&0,a_{j}^{2}=b_{j}^{2}=1.
\end{eqnarray}%
Then the Majorana representation of the Hamiltonian is

\begin{eqnarray}
\mathcal{H} &=&-\frac{i}{4}\sum\limits_{j=1}^{N}\left[ \left( t+\Delta
\right) b_{j}a_{j+1}+\left( t-\Delta \right) b_{j+1}a_{j}\right]  \notag \\
&&-\frac{i}{2}\sum\limits_{j\neq 1,N/2+1}^{N}\mu a_{j}b_{j}+\text{\textrm{%
H.c.}}  \notag \\
&&+\frac{\gamma }{2}(a_{1}b_{1}-a_{N/2+1}b_{N/2+1}-\text{\textrm{H.c.}}).
\end{eqnarray}%
The diagonalization of $H$ is directly related to a model of two coupled SSH
chain, which was systematically studied in Ref. \cite{Li}.

\section{Edge modes}

\label{Edge modes} Majorana edge mode plays an important role in the
characterization of the topological feature of matter, such as the bulk-edge
correspondence. The Majorana particle is localized at two ends. Since the
pattern of the zero mode is exponentially decaying, the bound\ Majorana
particle must require infinite chain condition. Therefore the conventional
zero mode cannot exist in a Hermitian system except for some trivial cases
\cite{LSZG}. In this paper, we focus on a finite-size non-Hermitian system
and we define the edge mode by the evanescent wave characterization.

We consider a Kitaev ring at the symmetric point $\Delta =t$. We write down
the Hamiltonian as the form
\begin{equation}
\mathcal{H}=\psi ^{T}h\psi ,
\end{equation}%
in the basis $\psi ^{T}=(a_{1},$ $b_{1},$ $a_{2},$ $b_{2},$ $a_{3},$ $b_{3},$
$...,$ $a_{N},b_{N})$, where $h$\ represents a $2N\times 2N$ matrix. Here
matrix $h$\ is explicitly written as%
\begin{eqnarray}
h &=&-\frac{i}{2}(\sum_{l=1}^{N}t\left\vert 2l\right\rangle \left\langle
2l+1\right\vert +\sum\limits_{l\neq 1,N/2+1}^{N}\mu \left\vert
2l-1\right\rangle \left\langle 2l\right\vert +\text{\textrm{H.c.}})  \notag
\\
&&+\frac{\gamma }{2}(\left\vert 1\right\rangle \left\langle 2\right\vert
-\left\vert N+1\right\rangle \left\langle N+2\right\vert -\text{\textrm{H.c.}%
})
\end{eqnarray}%
where basis $\left\{ \left\vert j\right\rangle ,j\in \left[ 1,2N\right]
\right\} \ $is an orthonormal complete set, $\langle j\left\vert j^{\prime
}\right\rangle =\delta _{jj^{\prime }}$. By taking the linear transformation%
\begin{equation}
\left\{
\begin{array}{c}
\left\vert \sigma ,2l-1\right\rangle =\frac{e^{-i\pi /4}}{2}(\left\vert
2l\right\rangle +i\sigma \left\vert 2N+3-2l\right\rangle ) \\
\left\vert \sigma ,2l\right\rangle =\frac{e^{i\pi /4}}{2}(\left\vert
2l+1\right\rangle -i\sigma \left\vert 2N+2-2l\right\rangle )%
\end{array}%
\right.
\end{equation}%
with $l\in \left[ 1,N/2\right] $ and $\sigma =\pm $, we can express matrix $h
$\ as $h=h_{+}+h_{-}$ with
\begin{equation}
\lbrack h_{+},h_{-}]=0,
\end{equation}%
where%
\begin{eqnarray}
h_{\sigma } &=&t\sum\limits_{l=1}^{N/2}\left\vert \sigma ,2l-1\right\rangle
\left\langle \sigma ,2l\right\vert -\mu \sum\limits_{j=1}^{N/2-1}\left\vert
\sigma ,2l\right\rangle \left\langle \sigma ,2l+1\right\vert   \notag \\
&&+\text{\textrm{H.c.}}+\sigma i\gamma (\left\vert \sigma ,1\right\rangle
\left\langle \sigma ,1\right\vert -\left\vert \sigma ,N\right\rangle
\left\langle \sigma ,N\right\vert ).
\end{eqnarray}%
Obviously, $h_{\pm }$\ describes two identical SSH chains with opposite
imaginary ending potentials $\pm i\gamma $. It is a $\mathcal{PT}$ symmetric
system and has been studied in Refs. \cite{LS} for $\mu >t$ and the
topological phase in the similar systems has been studied in Refs. \cite%
{Klett}. It is shown that there is a single zero mode in such a finite
degree of non-Hermiticity at the exceptional point (EP).

Now we concentrate on a single non-Hermitian SSH chain with the Hamiltonian%
\begin{eqnarray}
h_{\mathrm{SSH}} &=&\sum_{l=1}^{N/2}\left\vert 2l-1\right\rangle
\left\langle 2l\right\vert -\sum_{l=1}^{N/2-1}\mu \left\vert 2l\right\rangle
\left\langle 2l+1\right\vert   \notag \\
&&+\text{\textrm{H.c.}}+i\gamma (\left\vert 1\right\rangle \left\langle
1\right\vert -\left\vert N\right\rangle \left\langle N\right\vert ),
\label{H_SSH}
\end{eqnarray}%
where we take $t=1$\ and $\mu >0$\ for the sake of simplicity. Note that $h_{%
\mathrm{SSH}}$\ and $h_{\mathrm{SSH}}^{\dagger }$\ represent $h_{+}$\ and $%
h_{-}$, respectively. Here we redefine the space-reflection operator, or the
parity operator $\mathcal{P}$\ in the space spanned by the complete set $%
\{\left\vert l\right\rangle \}$. The action $\mathcal{P}$ on $\left\vert
l\right\rangle $ is given by the equality%
\begin{equation}
\mathcal{P}\left\vert l\right\rangle =\left\vert N+1-l\right\rangle .
\end{equation}%
Then we have $[h_{\mathrm{SSH}},\mathcal{PT}]=0$, which implies that $h_{%
\mathrm{SSH}}$\ is pseudo-Hermitian \cite{MostafazadehJPA}, i.e., $h_{%
\mathrm{SSH}}$ has either real spectrum or its complex eigenvalues occur in
complex conjugate pairs. On the transition from a pair of real levels to a
complex conjugate pair, or EP, two levels coalesce to a single level. The
key point of our approach is to link the zero-eigenvalue coalescing
eigenvector to the conventional Majorana zero mode in the thermodynamic
limit.

The Bethe ansatz wave function $\left\vert \psi _{k}\right\rangle
=\sum_{l=1}^{N}f_{l}^{k}\left\vert l\right\rangle $ has the form%
\begin{equation}
f_{l}^{k}=\left\{
\begin{array}{cc}
A_{k}e^{ikl}+B_{k}e^{-ikl}, & l=2j-1 \\
C_{k}e^{ikl}+D_{k}e^{-ikl}, & l=2j%
\end{array}%
\right. ,
\end{equation}%
where $j=1,2,\cdots ,$ $N/2$. Following the derivation in the Appendix A,
when the imaginary potential takes the value
\begin{equation}
\gamma =\mu ^{1-N/2},
\end{equation}%
there are three types of eigenvector $\left\vert \psi _{k}\right\rangle $,
with the eigenvalue%
\begin{equation}
\varepsilon _{k}=\pm \sqrt{1+\mu ^{2}-\mu \left( e^{2ik}+e^{-2ik}\right) }.
\label{energy}
\end{equation}%
In general, the EP varies as the system size changes. In the present model, $%
\gamma =\mu ^{1-N/2}$ is $N$ dependent except when $\mu =1$. It is
reasonable that the system is always at the EP for any $N$. We note that in
the case of $\mu =1$, it reduces to a uniform chain with $\gamma =1$, which
was studied in Ref. \cite{JLPT}. The solutions for $\mu \neq 1$\ is
concluded as follows.

(i) Scattering vector with real eigenvalues: In this case, $k$ is a real
number, the eigenvalue is real. The energy gap has a lower bound%
\begin{equation}
\Delta _{\mathrm{Gap}}\geq 2\left\vert 1-\mu \right\vert ,
\end{equation}%
which is crucial to protect the degenerate ground states of the original
Kitaev model from decoherence.

(ii) Coalescing vector with zero eigenvalue: Here the wave vector is
imaginary, $k=$ $\pm \frac{i}{2}\ln \mu $. The eigenvector has the form%
\begin{equation}
\left\vert \psi _{\mathrm{zm}}\right\rangle =\Omega \sum_{j=1}^{N/2}(\mu
^{1-j}\left\vert 2j-1\right\rangle -i\mu ^{j-N/2}\left\vert 2j\right\rangle
),  \label{WF_ZM}
\end{equation}%
satisfying%
\begin{equation}
h_{\mathrm{SSH}}\left\vert \psi _{\mathrm{zm}}\right\rangle =0.
\end{equation}%
where $\Omega =\mu ^{N/2-1}\sqrt{\left( 1-\mu ^{2}\right) /\left( 2-2\mu
^{N}\right) }$ is the Dirac normalizing constant. It is $\mathcal{PT}$
symmetric and has a zero biorthogonal norm, indicating the coalescence of
two levels. Actually, the zero-mode vector for $h_{\mathrm{SSH}}^{\dagger }$
can be constructed as%
\begin{equation}
\left\vert \eta _{\mathrm{zm}}\right\rangle =\Omega \sum_{j=1}^{N/2}(\mu
^{1-j}\left\vert 2j-1\right\rangle +i\mu ^{j-N/2}\left\vert 2j\right\rangle
),
\end{equation}%
satisfying
\begin{equation}
h_{\mathrm{SSH}}^{\dagger }\left\vert \eta _{\mathrm{zm}}\right\rangle =0.
\end{equation}%
On the other hand, it is easy to check
\begin{equation}
\langle \eta _{\mathrm{zm}}\left\vert \psi _{\mathrm{zm}}\right\rangle =0,
\label{zero norm}
\end{equation}%
and

\begin{equation}
\left\vert \eta _{\mathrm{zm}}\right\rangle =i\mathcal{P}\left\vert \psi _{%
\mathrm{zm}}\right\rangle \text{ or }\left\vert \eta _{\mathrm{zm}%
}\right\rangle =\mathcal{(}\left\vert \psi _{\mathrm{zm}}\right\rangle
)^{\ast },
\end{equation}%
which indicate that the vector $\left\vert \psi _{\mathrm{zm}}\right\rangle $%
\ has a zero biorthogonal norm and the relations between two conjugate
vectors. Based on these facts we conclude that the zero-mode vector $%
\left\vert \psi _{\mathrm{zm}}\right\rangle $\ is a coalescing vector for $%
\mu \neq 1$. The exact wave function of $\left\vert \psi _{\mathrm{zm}%
}\right\rangle $\ clearly indicates that it is a\ superposition of two parts
with nonzero amplitudes only located at even or odd sites, respectively. We
will show that vectors $\left\vert \eta _{\mathrm{zm}}\right\rangle $ and $%
\left\vert \psi _{\mathrm{zm}}\right\rangle $ have a close relation to the
standard\ zero modes of a Hermitian chain in the thermodynamic limit.

(iii) Evanescent wave vector with imaginary eigenvalue: The derivation in
the Appendix A shows that this kind of state only appears at $\mu <1$. In
this case, $k$ is still imaginary. Since matrix $h_{\mathrm{SSH}}$\ is
pseudo-Hermitian, this type of eigenvector always appears in pair. The\ wave
vector approximately takes

\begin{equation}
k=\pm i\frac{N-1}{2}\ln \mu
\end{equation}%
in large $N$ or small $\mu $\ and the wave function reads%
\begin{equation}
\left\vert \psi _{\mathrm{IM}}^{\sigma }\right\rangle \approx \frac{1}{2}%
\left[ \left( 1+\sigma \right) \left\vert 1\right\rangle +\left( 1-\sigma
\right) \left\vert N\right\rangle \right] ,
\end{equation}%
satisfying%
\begin{equation}
h_{\mathrm{SSH}}\left\vert \psi _{\mathrm{IM}}^{\sigma }\right\rangle
=i\sigma \left\vert \varepsilon _{\mathrm{IM}}\right\vert \left\vert \psi _{%
\mathrm{IM}}^{\sigma }\right\rangle .
\end{equation}%
with approximate eigenvalue%
\begin{equation}
\varepsilon _{\mathrm{IM}}\approx \pm i\mu ^{1-N/2},
\end{equation}%
We can see that the $\mathcal{PT}$ symmetry is broken. Both the two vectors
present evanescent wave and
\begin{equation}
\left\vert \psi _{\mathrm{IM}}^{\sigma }\right\rangle =\mathcal{PT}%
\left\vert \psi _{\mathrm{IM}}^{-\sigma }\right\rangle ,
\end{equation}%
which indicates that $\left\vert \psi _{\mathrm{IM}}^{\pm }\right\rangle $
are symmetry breaking. We summarize the solutions in Table \ref{Table I}.

\begin{table}[tbp]
\caption{For $N$-site system, $\protect\gamma =\protect\mu ^{1-N/2},\ n_{\mathrm{I}}$\
is the number of imaginary levels, $n_{\mathrm{EP}}\ $is the number of
coalescing state, $n_{\mathrm{S}}$ is the number of real levels.
We have $n_{\mathrm{I}}+2n_{\mathrm{EP}}+n_{\mathrm{S}}=N$.} \label{Table I}%
\renewcommand\arraystretch{1}

\begin{tabular}{p{1.2cm}p{1cm}p{1cm}p{1cm}}
\hline\hline
$\mu$ & $n_{\mathrm{I}}$ & $n_{\mathrm{EP}}$ & $n_{\mathrm{S}}$ \\ \hline
$\mu>1$ & $0$ & $1$ & $N-2$ \\
$\mu<1$ & $2$ & $1$ & $N-4$ \\ \hline\hline
\end{tabular}
\end{table}

It is well known that the number of Majorana zero mode for an infinite
Hermitian Kitaev chain with open boundary conditions can be a topological
invariant, referred as the bulk-edge correspondence. Similarly, the result
in Table I indicates the number of Majorana edge modes\ in a finite
non-Hermitian system can also be the topological invariant to identify the
topological phase of the corresponding bulk Hermitian system. This can be
regarded as a variant of the bulk-edge correspondence in the complex regime.
Recently, non-Hermitian SSH chains were experimentally realized by coupled
dielectric microwave resonators \cite{Scho,Poli} and photonic lattices \cite%
{Zeuner,Weim}. In Appendix B, we provide exact solutions for a $6$-site
system to demonstrate our main idea, which can be a protocol for the
experimental investigation.

\section{Connection to conventional zero mode}

\label{Connection to conventional zero mode}

In this section, we investigate the connection for the zero-mode states
between the present non-Hermitian model and infinite Hermitian SSH chain. We
consider $h_{\mathrm{SSH}}$\ in the large $N$ limit and analyze the
solutions in the following two regions.

(i) In the case of $\mu >1$, we have $\gamma =\mu ^{1-N/2}\rightarrow 0$.
Matrices $h_{\mathrm{SSH}}$\ and $h_{\mathrm{SSH}}^{\dag }$\ become the same
matrix of $N$-site single-particle Hermitian SSH chain with open boundary
conditions. Two zero-mode wave functions $\left\vert \eta _{\mathrm{zm}%
}\right\rangle $ and $\left\vert \psi _{\mathrm{zm}}\right\rangle $ become
two degenerate zero modes of the same Hermitian SSH chain. Remarkably, the
Eq. (\ref{zero norm}) for characterizing the coalescing levels is nothing
but the Dirac orthogonality of two degenerate zero modes. On the other hand,
we note that the amplitudes in $\left\vert \eta _{\mathrm{zm}}\right\rangle $
and $\left\vert \psi _{\mathrm{zm}}\right\rangle $ are only determined by $%
\mu $, independent of $N$. The coalescing zero mode is the finite-size
projection of the conventional degenerate zero modes in a Hermitian infinite
system with the open boundary condition.\ In this sense, the coalescing zero
modes $\left\vert \eta _{\mathrm{zm}}\right\rangle $ and $\left\vert \psi _{%
\mathrm{zm}}\right\rangle $ for any small $N$ carry the complete information
of conventional zero modes for the SSH chain.

(ii) Now we turn to the case of $\mu <1$. In contrast to the case (i), we
have $\gamma =\mu ^{1-N/2}$ $\gg 1$. Therefore, two ending sites are
adiabatically eliminated from the $N$-site chain. The original system is
separated into three independent parts. Two ending-site parts possess two
eigenvectors with imaginary eigenvalues $\pm i\gamma $. The third part
corresponds to an $(N-2)$-site single-particle Hermitian SSH chain with open
boundary conditions. Although an SSH chain is at total different\
topological phases for $\mu >1$ and $\mu <1$, respectively. However, such
an\ $(N-2)$-site Hermitian SSH chain\ is the same as that of the $N$-site
one in the large $N$ limit. In this sense, the coalescing zero mode for $\mu
<1$\ represents the same feature as that of $\mu >1$.

We plot Dirac norm distribution of the coalescing zero mode\ from Eq. (\ref%
{WF_ZM}) for the Hamiltonian (\ref{H_SSH}) with $N=30$, $22$, and $14$,
where $P\left( j\right) =\vert \left\langle j\right\vert \psi _{\mathrm{zm}%
}\rangle\vert ,j=1,2,3,\ldots ,N$, as the demonstration of our main result.
We see that systems with different size share a common part of wave vector.\
It indicates that one can retrieve the information of the Majorana zero mode
in the thermodynamic limit from a small non-Hermitian system. Two degenerate
conventional zero modes corresponds to the left and right vectors of the
coalescing zero mode.

\begin{figure}[tbp]
\includegraphics[ bb=71 228 449 551, width=0.45\textwidth, clip]{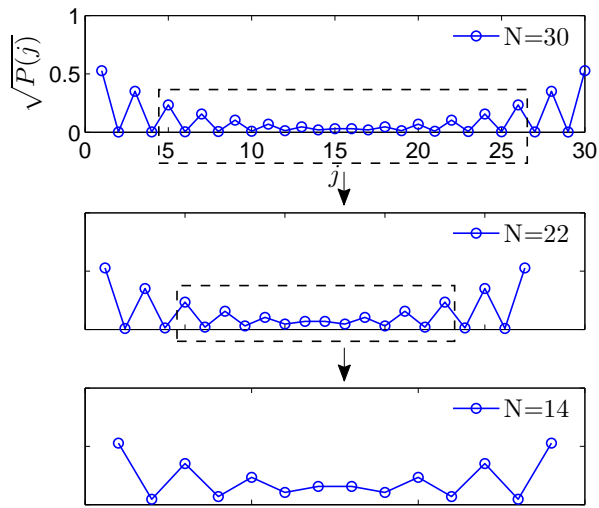}
\caption{(Color online) Plots of Dirac norm distribution of the coalescing
zero mode from Eq. (\protect\ref{WF_ZM}) for the Hamiltonian (\protect\ref{H_SSH}) with $\protect\mu =1.5,$\ $\protect\gamma =\protect\mu ^{1-N/2}$\
for different $N$. The dashed boxes indicate that the plot for $N=22$ is a
part of $N=30$, while the plot of $N=14$ is a part of $N=22$. Three systems
share a common part of wave vector.} \label{fig2}
\end{figure}

\section{Summary}

\label{Summary}

In conclusion, we have demonstrated a connection between the topological
characterization of an infinite system and the property of a specific small
system through a concrete Kitaev model. We have shown that a fine-tuned
non-Hermitian $\mathcal{PT}$-symmetric chemical potential in the finite-size
Kitaev ring results in coalescing Majorana zero\ mode. In particular, such a
zero mode in small lattice contains all the information of the conventional
degenerate Majorana zero modes in the thermodynamic limit. In addition, the
number of edge modes, which includes a pair modes with imaginary
eigenvalues, can be a topological invariant to characterize the quantum
phase diagram of the corresponding bulk Hermitian system. Although the
obtained conclusion is only based on a specific model, it may be a universal
feature of the topological system. The underlying mechanism is two fold.
Firstly, a single unit cell contains all the information about the bulk
system, including the topological feature. Secondly, a finite-size
non-Hermitian system, in some way, can be regarded as a sub-system embedded
in an infinite system.

\appendix*

\section{Evanescent solutions for non-Hermitian SSH chain}

\subsection{A. General solution}

In this Appendix, we provide the evanescent solutions of the Hamiltonian (%
\ref{H_SSH}) with $\gamma =\mu ^{1-N/2}$ by the Bethe ansatz method. The
Bethe ansatz wave function has the form%
\begin{equation}
f_{l}^{k}=\left\{
\begin{array}{cc}
A_{k}e^{ikl}+B_{k}e^{-ikl}, & l=2j-1 \\
C_{k}e^{ikl}+D_{k}e^{-ikl}, & l=2j%
\end{array}%
\right. ,
\end{equation}%
where $j=1,2,\cdots ,N/2$. The explicit form of Schrodinger equation
\begin{equation}
H\left\vert \psi _{k}\right\rangle =\varepsilon _{k}\left\vert \psi
_{k}\right\rangle
\end{equation}%
is expressed as%
\begin{equation}
\left\{
\begin{array}{c}
f_{m-1}^{k}-\mu f_{m+1}^{k}=\varepsilon _{k}f_{m}^{k} \\
f_{m}^{k}-\mu f_{m-2}^{k}=\varepsilon _{k}f_{m-1}^{k} \\
f_{m+2}^{k}-\mu f_{m}^{k}=\varepsilon _{k}f_{m+1}^{k} \\
f_{m+1}^{k}-\mu f_{m+3}^{k}=\varepsilon _{k}f_{m+2}^{k}%
\end{array}%
\right.  \label{bulk}
\end{equation}%
in the bulk and%
\begin{equation}
\left\{
\begin{array}{c}
i\gamma f_{1}^{k}+f_{2}^{k}=\varepsilon _{k}f_{1}^{k} \\
f_{N-1}^{k}-i\gamma f_{N}^{k}=\varepsilon _{k}f_{N}^{k}%
\end{array}%
\right.  \label{end}
\end{equation}%
at two ends, where $m=2j$, $j=1,2,3,\cdots ,N/2$. From Eq. (\ref{bulk}) we
have the spectrum%
\begin{equation}
\varepsilon _{k}=\pm \sqrt{1+\mu ^{2}-\mu \left( e^{2ik}+e^{-2ik}\right) },
\label{spectrum}
\end{equation}%
and coefficients%
\begin{equation}
\frac{B_{k}}{D_{k}}=\frac{C_{k}}{A_{k}}=e^{-ik}\sqrt{\frac{1-\mu e^{2ik}}{%
1-\mu e^{-2ik}}}.  \label{ABCD}
\end{equation}%
Together with Eq. (\ref{end}), we get the equation about $k$%
\begin{eqnarray}
&&\left( \varepsilon _{k}^{2}-\gamma ^{2}-1\right) [e^{i\left( N-2\right)
k}-e^{-i\left( N-2\right) k}]  \notag \\
+\mu \lbrack e^{i\left( N-4\right) k} &&-e^{-i\left( N-4\right) k}]+\mu
\left( \gamma ^{2}+\varepsilon _{k}^{2}\right) \left( e^{ikN}-e^{-ikN}\right)
\notag \\
=0. &&  \label{T eq1}
\end{eqnarray}%
In general, the wave function $\left\vert \psi _{k}\right\rangle $ with real
$k$ always represents the scattering vector. Since the real $k$ is bounded
by $\pm \pi $, the gap between two branches of eigenvalues $\varepsilon _{k}$%
\ is also bounded by $2\left\vert 1-\mu \right\vert $.

We are interested in the evanescent wave solution which corresponds to $%
k=i\kappa $\ or $\pi +i\kappa $ ($\kappa $\ is a real number). In this case,
Eq. (\ref{T eq1}) becomes%
\begin{eqnarray}
&&\left( \varepsilon _{k}^{2}-\gamma ^{2}-1\right) \sinh [\left( N-2\right)
\kappa ]+\mu \sinh [\left( N-4\right) \kappa ]  \notag \\
&&+\mu \left( \gamma ^{2}+\varepsilon _{k}^{2}\right) \sinh (N\kappa )=0,
\label{T eq2}
\end{eqnarray}%
with eigenvalue%
\begin{equation}
\varepsilon _{k}=\pm \sqrt{1+\mu ^{2}-2\mu \cosh (2\kappa )}.
\end{equation}%
At first, it is not hard to find that there are always two solutions
\begin{equation}
\kappa =\pm \frac{1}{2}\ln \mu
\end{equation}%
for\ Eq. (\ref{T eq2}). It leads to $A_{k}=D_{k}=0$ with $\kappa =-\frac{1}{2%
}\ln \mu $ (or $B_{k}=C_{k}=0$ with $\kappa =\frac{1}{2}\ln \mu $) according
to Eq. (\ref{ABCD}). This solution holds for all value of $\mu \neq 1$.

Secondly, in the case of
\begin{equation}
e^{-\kappa N},e^{-\kappa (N-2)},e^{-\kappa (N-4)}\ll 1,  \label{condition}
\end{equation}%
Eq. (\ref{T eq2}) can be reduced to%
\begin{eqnarray}
\left( \varepsilon _{k}^{2}-\gamma ^{2}-1\right) e^{\kappa (N-2)}+\mu
e^{\kappa (N-4)} &&  \notag \\
+\mu \left( \gamma ^{2}+\varepsilon _{k}^{2}\right) e^{\kappa N}=0, &&
\end{eqnarray}%
or a more popular\ form%
\begin{equation}
e^{-4\kappa }+\frac{\varepsilon _{k}^{2}-\gamma ^{2}-1}{\mu }e^{-2\kappa
}+\left( \gamma ^{2}+\varepsilon _{k}^{2}\right) =0.
\end{equation}%
Submitting the expression of $\varepsilon _{k}$ into the equation above, it
can be approximately reduced to a linear equation for $e^{-2\kappa }$, which
has the solution%
\begin{equation}
\kappa =\frac{1-N}{2}\ln \mu .
\end{equation}%
We note that the relations%
\begin{equation}
\left\{
\begin{array}{c}
\kappa _{\mu =1}=0 \\
\frac{\partial \kappa }{\partial \mu }<0%
\end{array}%
\right. ,
\end{equation}%
ensure the condition in Eq. (\ref{condition}) can be satisfied for the
region $\mu <1$. Then the obtained solution is justified. A similar
procedure can be performed in the case of $e^{\kappa N},e^{\kappa
(N-2)},e^{\kappa (N-4)}\ll 1$. In summary, a pair of solutions with
imaginary wave vectors are

\begin{equation}
k=\pm i\frac{1-N}{2}\ln \mu .
\end{equation}%
And the corresponding eigenvalues are%
\begin{equation}
\varepsilon _{\mathrm{IM}}\approx \pm i\mu ^{1-N/2}.
\end{equation}

\subsection{B. Example solution for $N=6$}

We demonstrate the above analysis via exact solutions for $N=6$ system.
Taking $\mu =2$, we have $\gamma =1/4$, matrix $h_{\mathrm{SSH}}$ is
expressed explicitly as

\begin{equation}
M_{1}=\left(
\begin{array}{cccccc}
i/4 & 1 & 0 & 0 & 0 & 0 \\
1 & 0 & 2 & 0 & 0 & 0 \\
0 & 2 & 0 & 1 & 0 & 0 \\
0 & 0 & 1 & 0 & 2 & 0 \\
0 & 0 & 0 & 2 & 0 & 1 \\
0 & 0 & 0 & 0 & 1 & -i/4%
\end{array}%
\right) .
\end{equation}%
The eigenvalues are%
\begin{eqnarray}
\varepsilon _{1} &=&\varepsilon _{2}=0,  \notag \\
\varepsilon _{3} &=&-\varepsilon _{4}=\frac{1}{8}\sqrt{350+2\sqrt{3553}}, \\
\varepsilon _{5} &=&-\varepsilon _{6}=\frac{1}{8}\sqrt{350-2\sqrt{3553}},
\notag
\end{eqnarray}%
which are all real. Two zero-mode eigen vectors are identical, i.e.,
\begin{equation}
\phi _{1}=\phi _{2}=(4i,1,-2i,-2,i,4),
\end{equation}%
which has zero biorthogonal norm.

Taking $\mu =1/2$, we have $\gamma =4$, matrix $h_{\mathrm{SSH}}$ is
expressed explicitly as

\begin{equation}
M_{2}=\left(
\begin{array}{cccccc}
i4 & 1 & 0 & 0 & 0 & 0 \\
1 & 0 & 1/2 & 0 & 0 & 0 \\
0 & 1/2 & 0 & 1 & 0 & 0 \\
0 & 0 & 1 & 0 & 1/2 & 0 \\
0 & 0 & 0 & 1/2 & 0 & 1 \\
0 & 0 & 0 & 0 & 1 & -i4%
\end{array}%
\right) ,
\end{equation}%
The eigenvalues are%
\begin{eqnarray}
\varepsilon _{1} &=&\varepsilon _{2}=0,  \notag \\
\varepsilon _{3} &=&-\varepsilon _{4}=\frac{i}{2}\sqrt{2\sqrt{238}+25}, \\
\varepsilon _{5} &=&-\varepsilon _{6}=\frac{1}{2}\sqrt{2\sqrt{238}-25},
\notag
\end{eqnarray}%
which contains a pair of imaginary numbers. Two zero-mode eigen vectors are
identical, i.e.
\begin{equation}
\phi _{1}=\phi _{2}=(i,4,-2i,-2,4i,1),
\end{equation}%
which has zero biorthogonal norm. In both two cases, vectors $\phi _{1},\phi
_{2}$\ are not normalized.\

\acknowledgments We acknowledge the support of the CNSF (Grant No. 11374163).

\end{document}